\documentclass[fleqn,usenatbib]{mnras}

\usepackage{newtxtext,newtxmath}
\usepackage[T1]{fontenc}
\usepackage{ae,aecompl}

\usepackage{graphicx}   % Including figure files
\usepackage{amsmath}   % Advanced maths commands
\usepackage{amssymb}   % Extra maths symbols

\newcommand{\fAGN}{f_{\rm AGN}}
\newcommand{\fagn}{f_{\rm AGN}}
\newcommand{\fSFG}{f_{\rm SFG}}

\newcommand{\rtwenty}{\rho_{20}}

\newcommand{\ana}{A\&A}

\newcommand{\kms}{{\rm km~s^{-1}}}
\newcommand{\kpc}{{\rm h^{-1}~kpc}}

\newcommand{\SFRf}{\rm SFR_{fib}}

\newcommand{\SFRt}{\rm SFR_{tot}}
\newcommand{\SFRfib}{\rm SFR_{fib}}

\newcommand{\barrho}{\bar{\rho}}
\newcommand{\rpvir}{r_{\rm p}/r_{\rm vir,nei}}
\newcommand{\rp}{r_{\rm p}}
\newcommand{\rvir}{r_{\rm vir, nei}}

\title[Environmental effect on AGN in Spiral galaxy]
{Direct effects of the environment on AGN triggering in SDSS spiral galaxies: merger-AGN connection}

\author[M. Kim et al.]{
Minbae Kim,$^{1}$\thanks{E-mail: mbkim@khu.ac.kr}
Yun-Young Choi,$^{1,2}$\thanks{E-mail: yy.choi@khu.ac.kr}
and Sungsoo S. Kim$^{1,2}$
\\
$^{1}$School of Space Research, Kyung Hee University, Yongin, Gyeonggi 17104, Republic of Korea \\
$^{2}$Department of Astronomy \& Space Science, Kyung Hee University, Yongin, Gyeonggi 17104, Republic of Korea
}

\date{Accepted XXX. Received YYY; in original form ZZZ}

\pubyear{2019}

\begin{document}
\label{firstpage}
\pagerange{\pageref{firstpage}--\pageref{lastpage}}
\maketitle

\begin{abstract}
We examine whether galaxy environments directly affect triggering nuclear activity 
in Sloan Digital Sky Survey (SDSS) local spiral galaxies using a volume-limited sample 
with the $r$-band absolute magnitude $M_{r} < -19.0$ and $0.02 < z < 0.055$ 
selected from the SDSS Data Release 7. 
To avoid incompleteness of the central velocity dispersion $\sigma$ of the volume-limited sample 
and to fix the black hole mass affecting AGN activity, 
we limit the sample to a narrow $\sigma$ range of $130~\kms<\sigma<200~\kms$. 
We define a variety of environments as a combination of neighbour interactions and local density on a galaxy. 
After the central star formation rate (which is closely related to AGN activity level) 
is additionally restricted, the direct impact of the environment is unveiled. 
In the outskirts of rich clusters, red spiral galaxies show a significant excess 
of the AGN fraction despite the lack of central gas. 
We argue that they have been pre-processed before entering the rich clusters, 
and due to mergers or strong encounters in the in-fall region, their remaining 
gases efficiently lose angular momentum. We investigate an environment 
in which  many star-forming galaxies coexist with a few starburst-AGN composite 
hosts having the highest [OIII] luminosity. We claim that they are 
a gas-rich merger product in groups or are group galaxies in-falling into clusters, 
indicating that many AGN signatures may be obscured following the merger events. 
\end{abstract}


%%%%%%%%%%%%%%%%%%%%%%%%%%%%%%%%%%%%%%%%%%%%%%%%%%

%%%%%%%%%%%%%%%%% BODY OF PAPER %%%%%%%%%%%%%%%%%%

\section{Introduction}
\label{sec:sec1}
It is generally accepted that 
the active galactic nuclei (AGNs) are powered 
by feeding supermassive black holes (SMBHs) 
via matter accretion at a galactic central region \citep{Lyndenbell1969, Rees1984}. 
AGN activity depends primarily on the availability of cold gas fuel and
the BH mass of their host galaxies, which are closely related to the properties of the host galaxy 
\citep{Moles1995, Choi2009, Schawinski2010, Sabater2012, Sabater2015, Argudo2018}, 
as well as to the gas transporting mechanisms to the galactic central region near the SMBHs.
Possible SMBHs feeding gas inflow mechanisms include internal dynamical processes such as 
bar-driven gas inflow \citep{Shlosman1990, Combes2003, Kormendy2004}, 
turbulence in interstellar matter \citep{Elmegreen1998, Wada2004, Wada2009} 
and stellar wind \citep{Ciotti2007, Davies2012}, 
as well as the external processes such as galaxy interactions and mergers 
\citep{Sanders1988, Hernquist1995, Dimatteo2005, Springel2005, Hopkins2006, Alonso2007, Hann2009}. 

Due to the very large data sets of galaxies provided by Sloan Digital Sky Survey (SDSS), 
statistically significant studies on the possible link between AGN activity and the environments of host galaxies
have become possible. 
Several studies have determined the correlation between galaxy interactions and AGN activity 
\citep{Ellison2011, Hwang2012, Liu2012, Satyapal2014, Weston2017, Goulding2018}. 
For example, \citet{Ellison2011} and \citet{Satyapal2014} found that the AGN fraction increases at small projected separations using SDSS galaxy samples of close pairs. 
\citet{Hwang2012} found that the environmental dependence of the AGN fraction varies with the host galaxy morphology.
\citet{Liu2012} also found that both AGN activity and recent star formation of host galaxies increase 
when pair separation decreases in AGN pair samples. 
Observations have also shown the role of the local density environment in AGN activity. 
\citet{vonderlinden2010} found 
that the AGN fraction of quiescent red galaxies in SDSS cluster galaxy samples 
decreases towards the centre of the clusters. 
\citet{Sabater2013} showed that the AGN fraction decreases towards denser environments 
using a density estimator defined using the distance to the 10th nearest neighbour. 

However, some observational studies have found contradictory results 
\citep{Grogin2005, Darg2010, Slav2011, Scott2014, Vill2014}. 
\citet{Darg2010} found no significant differences in the environmental distributions 
of merger samples and randomly selected control samples from SDSS and Galaxy Zoo samples. 
\citet{Slav2011} showed that Seyfert nuclei triggering is not directly related to the large-scale morphology 
or local environment of their host galaxies, including bars, rings and close companions. 
\citet{Grogin2005} showed that symmetry parameter of AGN hosts
is not different from that of non-AGN hosts, suggesting
that there is no relation between galaxy interaction and AGN activity.

On the other hand, when the morphology of their host galaxies is fixed, 
more interesting results have been revealed.  
Using the SDSS cluster galaxy sample, 
\citet{Hwang2012} investigated environmental effects on AGN activity 
by comparing galaxies in the cluster and field. 
For spiral galaxies, the AGN fraction does not change much, 
beginning to decrease only near the cluster centre. 
Elliptical galaxies in cluster have obviously lower AGN fractions than those in the field. 
\citet{Sohn2013} also found similar results where 
the AGN fractions for elliptical galaxies significantly decrease with increasing galaxy number density,
whereas the fractions for spiral galaxies change very little, even in compact groups where galaxies may
experience frequent galaxy interactions. 
\citet{Alonso2013} found that for face-on spiral AGN host galaxies, 
the powerful AGN fraction increases in denser environments 
using a projected local density parameter defined by the fifth nearest neighbour. 
\citet{Argudo2016} found the environmental effects are different 
depending on the stellar mass of the galaxy. 
The AGN fractions of very massive and isolated galaxies increase when transitioning from void to denser environment regions,  
whereas those of galaxies with low-mass decrease. 

Here, a question arises about the direct and indirect effects of the external factors on the AGN triggering.
The external factors effectively control the gas supply to galaxies, but 
it is not clear whether they directly or indirectly affect AGN activity 
in the innermost part of a galaxy. 
Therefore, when investigating the direct connection between AGN activity and galaxy environment, 
which we aim to do in this work, galaxy properties closely related to gas availability need to be fixed.

Motivated by this, 
we select `bulge-dominated' spiral galaxies that more dominantly host AGNs and are located across more diverse environments, 
ranging from void to cluster environments. 
To examine the direct environmental impact, we additionally fix a central star formation (SF)
because AGN activity is better linked to the central SF than to galaxy-wide SF or $u-r$ colour 
\citep{Kauffmann2007, Lamassa2013,Kiminprep}.
We use a fibre star formation rate, $\SFRf$ as a proxy of the central SF.
To ensure that the diameter (3 arcsec) of the SDSS optical fibre covers the inner few kpc of the sample galaxy,
we fix the redshift upper limit of the SDSS volume-limited sample to be $z = 0.055$. 
At the median redshift of our sample ($z = 0.043$), 
the fibre subtends to $\sim 2.7$ kpc, which is similar to the average size of bulges in spiral galaxies 
\citep{Fisher2010, Cheung2015}. 
As a result, the $\SFRf$s in our sample well represent more confined central activity.

Our sample selection and definitions of the environments 
are described in Section~\ref{sec:sec2}. 
For each target galaxy, we measure two different environmental parameters: 
level of galaxy–galaxy interaction 
(the projected separation of the nearest pair galaxy from the target galaxy, $\rp$) 
and large-scale environment 
(the background density described by the 20 closest galaxies, $\rtwenty$). 
In Section~\ref{sec:sec3}, we investigate their combined effects on the triggering of AGN.
We additionally limit the central $\SFRf$ of the sample
to isolate the effect on only AGN activity.
Our discussion and summary are presented in Sections~\ref{sec:sec4} and \ref{sec:sec5}.   
Throughout this paper, 
the cosmological parameters are assumed for the $\Lambda$CDM cosmological model with 
the density parameters $\Omega_m = 0.27$ and $\Omega_{\Lambda} = 0.73$.

\section{Observational Data Sample}
\label{sec:sec2} % used for referring to this section from elsewhere

\subsection{Sloan Digital Sky Survey}
\label{sec:sec2.1}

We select a volume-limited spiral-galaxy sample with
an $r$-band absolute magnitude ${M_{r} < -19.5}+5{\rm log}h$ 
(hereafter we exclude the $+5{\rm log} h$ term in the absolute magnitude term) 
and redshift $0.02 < z < 0.055$ 
from Sloan Digital Sky Survey Data Release 7 
\citep[SDSS DR7;][]{Abazajian2009}.
We also use several Value-Added Galaxy Catalogues (VAGCs) for SDSS DR7 
for the physical parameters of the host galaxies. 

For the fibre and total star formation rates ($\SFRfib$ and $\SFRt$ respectively)
and the information of optical emission lines of late-type galaxies in our sample, 
we adopt spectroscopic parameters from the Max Planck Institute for Astrophysics and 
Johns Hopkins University (MPA/JHU) DR8 catalogue. 
For calculating the $\SFRfib$ of star-forming galaxies (SFGs), emission lines were used \citep{Brinchmann2004}, 
while those for the others types were calculated using fibre photometry \citep{Salim2007}. 
We adopt photometry parameters 
such as ${M_{r}}$, $u-r$ colour, $\Delta (g-i)$, and $c_{\rm in}$ 
from the Korea Institute for Advanced Study DR7 Value-Added Galaxy Catalogue 
\citep[KIAS DR7-VAGC;][]{Park2005, Choi2010}, 
which is complementary to 
the New York University Value-Added Galaxy Catalogue \citep[NYU VAGC;][]{Blanton2005} and 
MPA/JHU catalogue. 

The stellar velocity dispersion $\sigma$ is a good indicator of BH mass 
\citep[${\rm M}_{BH}$–$\sigma$ relation;][]{Ferrarese2000, Gebhardt2000, Tremaine2002, Gultekin2009, Batiste2017}. 
We set $70~\kms$ as a lower limit on $\sigma$ measurements  
due to the instrumental resolution of the SDSS spectrograph.
The $\sigma$ measurements adopted from NYU-VAGC are used only for
spectra with median signal-to-noise (S/N) per pixel $>10$ and
are corrected for the aperture of an SDSS spectroscopic fibre  \citep{Bernardi2003}.

\subsection{Morphology classification}
\label{sec:sec2.2}

Morphology classification is adopted from the KIAS DR7-VAGC. 
The galaxy morphology is divided into
early-type (ellipticals and lenticulars) and 
late-type (spirals and irregulars) 
based on their locations in two parameter space \citep{Park2005}: 
$u-r$ colour versus $g-i$ colour gradient $(\Delta (g-i))$ space 
and $u-r$ colour versus $i$-band concentration index $(c_{\rm in})$ space. 
$\Delta (g-i)$ is the difference in $g-i$ colours 
between the outside $0.5R_{\rm Pet} < R < R_{\rm Pet}$ and the inside in $R < 0.5R_{\rm Pet}$ of the galaxy, 
where $R_{\rm Pet}$ is the Petrosian radius in the $i$-band. 
If $\Delta (g-i)$ is positive/negative, 
this means there is a bluer/redder core and redder/bluer outside. 
The $c_{\rm in}$ is defined as the ratio of $R_{50}$ and $R_{90}$ 
which are the Petrosian radii with fluxes of 50\% and 90\% smaller in the $i$-band. 
Commonly, the $c_{\rm in}$ of an early-type galaxy 
is smaller than that of a late-type galaxy. 
We use 11 096 late-type galaxies with $\sigma > 70~\kms$ for this study. 

\subsection{AGN selection}
\label{sec:sec2.3}

We classify the types of galaxies based on 
Baldwin–Phillips–Terlevich (BPT) diagrams 
\citep{Baldwin1981, Veilleux1987}. 
For $\rm H\alpha$, $\rm H\beta$, $[\rm OIII]\lambda5007$, $[\rm NII]\lambda6584$ 
as measured in the MPA/JHU Catalogue \citep[cf.][]{Brinchmann2004}, 
several flux ratios of narrow emission lines 
with ${\rm S/N} \geq 6$  are used. 
Galaxies can be classified as 
SFGs, starburst-AGN composite galaxies, and 
pure AGNs (Seyfert and low-ionization nuclear emission-line regions, LINERs). 
Composite galaxies are located between the maximum starburst line \citep{Kewley2001} 
and the pure star-forming line \citep{Kauffmann2003}. 

In this study, we define AGNs as Type II AGNs 
including composite galaxies and pure AGNs 
using the criterion of \citet{Kewley2006}. 
In our AGN sample, 
we exclude potential Type ${\rm I}$ AGNs, 
which have a full width at half-maximum (FWHM) 
of an $\rm H\alpha$ emission line larger than $\sim 500~\kms$. 
% Retired galaxies
Here, one should note that all the LINERs identified by BPT diagram are not bona fide AGNs, 
some of the weak LINERs are retired galaxies powered by hot low-mass evolved stars 
rather than low-luminosity AGNs 
\citep{Stasinska2008, Cidfernandes2010, Cidfernandes2011, Melnick2013}. 
By adopting the criterion of \citet{Cidfernandes2011}, 
we exclude LINERs and pure AGNs classified as ambiguous with $W_\mathrm{H\alpha} < 3\AA$. 
The result of the spectral type classification is listed in Table~\ref{tab:tab1}.
Out of 11 096 spiral galaxies with $\sigma > 70~\kms$, 2942 AGN hosts are found. 

\begin{table}
   \centering
   \caption{Spectral type classification statistics and comparison of results 
with different emission line S/N cuts}
   \label{tab:tab1}
   \begin{tabular}{lccr} % four columns, alignment for each
      \hline
      \hline      
       Number (Fraction) & ${\rm S/N} \geq 6$ & ${\rm S/N} \geq 3$ \\
      \hline
         Total ($\sigma > 70~\kms$)  & 11 096  (1.00)  & 11 096 (1.00) \\
         SFG                         &   3451  (0.31)  &   4743 (0.43) \\
         Total AGN                   &   2942  (0.27)  &   3973 (0.36) \\
         ~~~~~~~~~~Composite         &   2209  (0.20)  &   3153 (0.28) \\
         ~~~~~~~~~~Pure AGN          &    733  (0.07)  &    820 (0.07) \\
      \hline
   \end{tabular}
\end{table}

We conservatively select AGNs using a signal-to-noise ratio of ${\rm S/N} \geq 6$  
instead of the commonly used ${\rm S/N} \geq 3$. 
Table~\ref{tab:tab1} also shows 
how the galaxy number of each spectral type varies depending on the signal-to-noise. 
Unlike SFGs and composite galaxies, which drastically decrease, 
no significant change is found in the number of pure AGNs without retired galaxies. 
We confirm that the overall results do not change much 
when ${\rm S/N} \geq 3$ is used instead of ${\rm S/N} \geq 6$ 
in the definition of an AGN. 

\subsection{Environmental parameters}
\label{sec:sec2.4}

To understand the environmental dependence of AGN fraction, 
two different environmental factors are considered in this study. 
One is a large-scale background density $\rho_{20}$ defined 
by the 20 closest galaxies of a target galaxy in the sample. 
It is just over a few Mpc in scale 
\citep[see section 2.5 of ][ for details]{Park2009a}. 
The other is the distance between a target galaxy and the pair galaxy. 
To define the estimates, 
we use a volume-limited galaxy sample with
the $r$-band absolute magnitude ${M_{r} < -19.0}$ and $0.02 < z < 0.055$.
The full detail of the estimation for methods $\rho_{20}/\bar{\rho}$ and $\rpvir$ 
are described in \citet{Park2008} and \citet{Park2009a}.

\subsubsection{Large-scale background density}
\label{sec:sec2.4.1}

The large-scale background density of a target galaxy is given by 
\begin{equation} 
\rho_{20}({\bf x})/{\bar\rho} = \sum_{i=1}^{20} \gamma_i L_i W_i(|{\bf x}_i - 
{\bf x}|)/{\bar\rho}, 
\end{equation} 
where the $\gamma_i$ is the mass-to-light ratio of a background galaxy
which is adopted to obtain the mass density described by 20 neighbouring galaxies.
Here, the ratio of dark halo virial mass for early- and late-type targets,
$\gamma$(early) $= 2\gamma$ (late) is only needed.
\citet{Park2009a} derived the ratio from the velocity dispersion of 
neighbouring galaxies around a target and a relation between the velocity dispersion and the virial
halo mass of $M_{\rm vir} \propto \sigma_{\Delta v}^{\beta},
$ where $\beta =2.5$. The mass is related to the galaxy plus dark halo system. 
The mean density of the Universe with total volume $V$ is obtained by 
${\bar\rho} = \sum_{\rm all} \gamma_i L_i /V$, 
and $W_i(\bf x)$ and $L_i$ are a smoothing weight function and 
the $r$-band luminosity of the closest 20 background galaxies of a target spiral galaxy respectively.

For estimating the large-scale background density, 
we adopt the spline-kernel weight $W(r)$ as the smooth weight function
\citep{Monaghan1985}.
According to \citet{Park2007}, the adaptive smoothing kernel
better preserves a property of the galaxy distribution than a kernel with a fixed smoothing length
and has the advantage that the S/N for estimating density is more uniform.
\citet{Park2008} showed that 
the 20 galaxies required within the adaptive smoothing volume is close to the 
smallest number yielding good local density estimates. 

\subsubsection{The nearest neighbour galaxy}
\label{sec:sec2.4.2}
The pair galaxy for a host galaxy is defined using 
the conditions of $r$-band absolute magnitude and radial velocity difference 
as that which is located closest to the target galaxy in the sky.
If a host galaxy has the $r$-band absolute magnitude $M_r$, 
the nearest neighbour galaxy for that host galaxy has an $r$-band 
absolute magnitude brighter than $M_r+\Delta M_r$ and a radial velocity difference 
less than $\Delta v$, making it the most influential neighbour.
We adopted $\Delta M_r = 0.5$ and $\Delta v=400\kms$
obtained by measuring the pairwise velocity difference between target galaxies and their neighbours 
\citep[see section 2.4 of ][]{Park2008}.

The pair separation of $\rp$ measures the impact of interactions with the most influential neighbour.
The virial radius of the nearest neighbour (i.e., pair galaxy) $r_{\rm vir, nei}$ is defined as $\rp$, 
where the mean mass density $\rho_n$ within the sphere with radius of $r_p$ is equal to 740 times 
the mean density of the Universe $\bar{\rho}$:
\begin{equation}\label{eq:Dss}
\rvir = (3 \gamma_{\rm nei} L_{\rm nei} /4\pi \bar{\rho}/740)^{1/3},
\end{equation}
where $\gamma_{\rm nei}$ and $L_{\rm nei}$ are the mass-to-light ratio 
and the $r$-band luminosity of the nearest neighbour galaxy, respectively.
According to this formulae, the virial radii of spiral galaxies with $M_r=-19.5$ corresponds to $210~\kpc$.

\section{Results}
\label{sec:sec3}

\subsection{Strong dependence of AGN fraction on galactic central SF}
\label{sec:sec3.1}

We begin this study by investigating 
how AGN activity varies depending on two central quantities: 
the velocity dispersion $\sigma$ and central star formation rate $\SFRf$.

\begin{figure*}
    \centering   
   \includegraphics[scale=1]{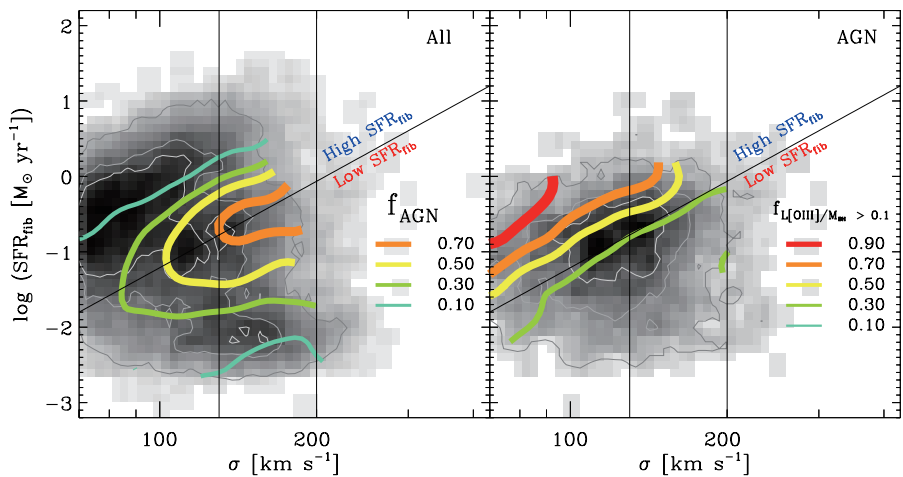}
    \caption{Distributions of the $\fagn$ in the sample galaxies (left) 
and of the powerful AGN fraction in AGN hosts (right) 
on the $\SFRf$–$\sigma$ plane.
A 2D histogram in each panel shows the number density distribution of galaxies (left) 
and of AGN hosts (right), respectively.
Each grey thin contour for the number density distribution encloses 
0.5$\sigma$, 1$\sigma$ and 2$\sigma$ in that order. 
For the $\fagn$ distribution, contours with an error greater than 30\% of the $\fagn$ measurement are eliminated.
The black solid diagonal line divides the sample into two subsamples of high-$\SFRf$ 
and low-$\SFRf$ samples, which are used in Section~\ref{sec:sec3.3}}.
    \label{fig:fig1}
\end{figure*}

Figure~\ref{fig:fig1} shows the distributions of the AGN fractions 
in the sample (in the left-hand panel) and of the powerful AGN fractions 
in all AGN hosts (in the right-hand panel) in the $\SFRf$-$\sigma$ space.
Each coloured thick-line contour denotes a constant fraction.
A two-dimensional histogram of the number density in each panel is plotted together
for all galaxies (in the left-hand panel) and AGN hosts (in the right-hand panel), respectively.
Superposed on the 2D histogram plot are the contours of constant number densities
enclosing 0.5$\sigma$, 1$\sigma$ and 2$\sigma$ in that order. 
All the smoothed distributions that we measure hereafter are obtained 
using the fixed-size spline kernel for each bin (60 by 60) 
in the parameter space.
The corresponding bin size for the $\SFRf$–$\sigma$ plane is 
$\Delta ({\rm log}{~\sigma})=0.013$ by $\Delta ({\rm log}{~\SFRf})=0.090$, 
and for the environment parameter space of $\rp$ and $\rtwenty$, 
the bin size is $\Delta ({\rm log}{~\rp/\rvir})=0.080$ 
by $\Delta ({\rm log}{~\rtwenty/\barrho})=0.080$. 

AGN activity is measured by calculating AGN occupation fraction $\fAGN$ 
as a ratio between AGN hosts and all galaxies. 
Assuming that the [OIII] emission-line luminosity $L_{\rm [OIII]}$ and 
$\sigma$ estimate the BH accretion rate and a BH mass, respectively, 
we calculate the [OIII] line luminosity normalized by BH mass, 
which measures how rapidly a BH has grown to its present-epoch mass 
(i.e., a tracer of AGN power).
The powerful AGN fraction is given by a ratio between AGNs with 
${\rm log}~L_{\rm [OIII]}/M_{\rm BH} > 0.1$ and all AGNs.
The $M_{\rm BH}$ is derived from the $M_{\rm BH}$-$\sigma$ relation 
for galaxies given by \citet{Tremaine2002}. 

Our sample galaxies show a trimodal number distribution 
in the $\SFRf$–$\sigma$ space that exhibits three distinct peaks: 
each peak is dominated by galaxies with three different spectral types, 
SFGs, AGN hosts, and quiescent galaxies.

At a given $\sigma$, AGNs in galaxies with the largest $\SFRf$s 
composing of an SFG sequence are 
rarely found but are the most powerful. 
$\fagn$ reaches its peak when both the $\SFRf$ (consisting of the SFG fraction, $\fSFG$) 
and the power of the detected AGN considerably decreases.
After the $\fagn$ peak, with decreasing $\SFRf$, 
$\fagn$ decreases and the reduced power of AGNs changes very little, 
i.e., AGNs seem to fade enough. 
The strong correlation between AGN luminosity and current SF 
is clearly revealed when measured over the central region, implying
that the $\SFRf$ (central SF activity) and $L_{\rm [OIII]}$ (BH accretion rate) 
are linked through the central gas supply.
For example, for the population with the highest $\SFRf$, 
$L_{\rm[OIII]}$ of their detected AGNs is also highest, but their $\fagn$ is rather lowest.
In other words, $\fagn$ is not directly connected with $\SFRf$ and $L_{\rm[OIII]}$.

Here we should keep in mind that the AGN fraction we calculated is 
not the probability of triggering AGN in a galaxy, 
but rather the probability of detecting the triggered AGN. 
This is because all triggered AGNs may not be detected on the optical band.
That is, a low value of $\fagn$ means that AGN is difficult to trigger or 
that it is difficult to detect the triggered AGN. 

Here, we assume that 
%the population with the highest $\SFRf$ and the lowest $\fagn$
if a large amount of gas funnels into the galactic centre, it drives starburst and a nuclear gas-inflow
and given the high $L_{\rm[OIII]}$ of the observed AGNs, 
many AGNs are triggered but obscured by dense gas and dust \citep[see][for a review]{Hickox2018}.
During this starburst phase, SMBHs may grow rapidly, raising their $\sigma$. 
The obscured AGNs are hidden in galaxies with higher $\SFRf$ and larger $\sigma$
that lie slightly above the SFG sequence \citep[e.g.,][]{Chang2017}. 
After galaxies leave the SFG sequence, 
many AGNs become suddenly observable at larger $\sigma$ and lower $\SFRf$
and their luminosity is no longer as luminous as before, 
implying that this process has taken place quite abruptly.

If so, how are they eventually unobscured? 
As a main cause of the rapid SF quenching, 
a strong gas outflow driven by powerful AGNs is usually 
invoked to eject the surrounding gas and dust \citep[e.g.,][]{Dimatteo2005, Hopkins2005}.
\citet{Ishibashi2016} claimed in this picture that AGN feedback and starburst are intrinsically coupled.
Accordingly, we argue that the population with the highest $\SFRf$ and the lowest $\fagn$ in our sample
favours the evolutionary picture of starburst-AGN connection.
Meanwhile, after most AGNs are observed,
the current SF depletes only remaining gases and the AGNs fade.

Because of the strong dependence of $\fagn$ on $\sigma$ and $\SFRf$,
it is not easy to observe whether environments directly affect  AGN activity
without restricting $\SFRf$ and $\sigma$ values of a sample.
For this purpose, we divide the sample into two according to the $\SFRf$ value 
of the AGN peak at a given $\sigma$.
A thin diagonal line in Figure~\ref{fig:fig1} separates the two subsamples.
We assume that each subsample has a different AGN mode (e.g. `powerful' and `weak' AGNs).

Next, let us note the strong $\sigma$ dependence of the AGN fraction. 
From the right-hand panel of Figure~\ref{fig:fig1}, 
when AGN power is given, the $L_{\rm [OIII]}$ of AGN hosts 
with a smaller $\sigma$ is also lower. 
For weak AGNs, 
detection of the [OIII] line in the smaller $\sigma$ galaxies is more difficult 
and the weak AGNs can be missed during AGN detection \citep[e.g.,][]{Aird2012}. 
Furthermore, our sample is missing many galaxies with a smaller $\sigma$ 
due to the cut of $M_r < -19.5$ adopted to select our volume-limited samples. 
Hereafter, to avoid the $L_{\rm [OIII]}$-related selection effects 
and incompleteness at lower $\sigma$ values, 
we only use a galaxy sample with $130~\kms<\sigma<200~\kms$. 
The BH mass range corresponding to the $\sigma$ range is 
$7.4 < \mathrm{log}(M_{\rm BH}/M_{\odot}) < 8.1$. 
The median and lower and upper quartile BH masses of this sample are 
$10^{7.56}$, $10^{7.44}$, and $10^{7.72} M_{\odot}$, respectively. 

\subsection{Environmental dependence of AGN fraction}
\label{sec:sec3.2}

\subsubsection{Galaxy interactions in diverse local environments}
\label{sec:sec3.2.1}
\begin{figure}
    \centering   
   \includegraphics[scale=1]{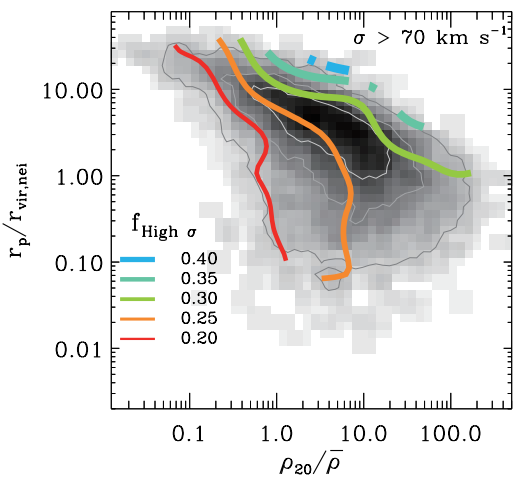}
    \caption{Distribution of the occupation fraction of galaxies with a high $\sigma$ of
 $130~\kms<\sigma<200~\kms$ for our sample in the $\rp$–$\rtwenty$ plane.
The coloured thick contour line denotes a constant fraction and 
the contours with an error greater than 30 percent of the fraction measurement are eliminated.
The grey 2D histogram represents 
the number density distributions of the sample and
the grey thin contours enclose 0.5$\sigma$, 1$\sigma$ and 2$\sigma$ of the sample, respectively.
}
    \label{fig:fig2}
\end{figure}

First, 
we limit the sample to a narrow $\sigma$ range of $130~\kms<\sigma<200~\kms$. 
Then, we examine how galaxy interactions and 
large-scale environments affect AGN activity by measuring
the dependence of AGN activity in the $\rp$ and $\rtwenty$ parameter space.
In Figure~\ref{fig:fig2}, 
we show the galaxy number distribution in the $\rp$-$\rtwenty$ space measured 
by using the total sample with $\sigma>70\kms$.

In fact, the measurements of the two environmental factors, $\rtwenty$ and $\rp$,
are entangled by their definitions. 
The background density estimate $\rtwenty$ spans various large-scale environments 
from voids to clusters.
For convenience, we arbitrary assign the very low density of $\rtwenty< \barrho$,  
intermediate density of $\barrho<\rtwenty<20\barrho$,
and high density of $\rtwenty>20\barrho$ to discrete environmental types such
as void/field, group, and cluster environments, respectively.
Here, recall that the background galaxies used to define 
the $\rtwenty$ are luminous ($M_r<-19.0$).
Meanwhile, the pair separation of $\rp$ measures the impact of interactions with the most influential neighbour.
According to \citet[][]{Park2009a}, when a galaxy approaches the virial radius of its neighbour ($\rp<\rvir$),
the two galaxies interact hydrodynamically, 
which is enough to change the mean morphology and 
SF activity of the target galaxies \citep[e.g.,][]{Park2009a}.
When $\rp$ decreases to about 0.05 times $\rvir$ (corresponding to $10 \sim 20 h^{-1} \rm kpc$ for our sample), 
the target and neighbour galaxies start to make physical contact with each other. 
At some point, the interacting galaxies eventually merge so that 
the second closest neighbour becomes the nearest neighbour, and consequently,
the end-product of the merger is more likely to be located at the largest $\rp$.

It should be noted that galaxies with $\rp > \rvir$ possess no influential neighbour nearby
but can be surrounded by less luminous background galaxies than the nearest neighbour.
We consider that among these, galaxies at $\rtwenty<\barrho$ where there exist few interacting pairs
with $\rp>\rvir$ are in a field environment that do not belong to any galaxy group or cluster.
These galaxies can be used as a control for understanding the effects of galaxy 
interactions and mergers in this study. 

The $\rp$–$\rtwenty$ diagram shows how these two parameters are intricately related.  
At $\rp>\rvir$, the $\rp$ and $\rtwenty$ are largely closely correlated, but  
at smaller pair separations below $\rp\sim0.5\rvir$, the correlation tends to disappear. 
Accordingly, the intermediate background density surrounding a target galaxy 
provides the widest $\rp$ distribution.
In other words, the $\rp$–$\rtwenty$ diagram demonstrates that a single measurement of $\rp$ or $\rtwenty$ 
cannot properly describe the environment in which a galaxy resides. 

Each coloured thick contour in Figure~\ref{fig:fig2} 
denotes a constant fraction of galaxies having $130\kms<\sigma<200\kms$.
The contours show the $\sigma$ distribution over the $\rp$–$\rtwenty$ space.
At the largest $\rp$s in the intermediate-$\rtwenty$ region, 
the highest occupation fraction of the high-$\sigma$ galaxies is found,  
suggesting that mergers contribute significantly to bulge growth.
The location provides a suitable environment for observing the merger-AGN connection 
that SMBH-galaxy co-evolution is driven by galaxy mergers
\citep[e.g.,][]{Dimatteo2005, Hopkins2008, Alexander2012, Treister2012}.

\begin{figure}
  \centering   
 \includegraphics[scale=1]{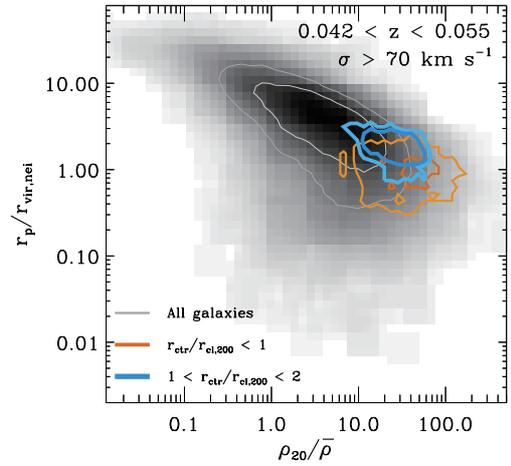}
 \caption{Distribution of cluster member galaxies in the galaxy sample
 with $0.042 \leq z <0.055$ and $\sigma > 70~\kms$. 
  The number distribution of the whole sample is represented by the grey 2D histogram and grey thin contours. 
 The orange and blue thin contours show the number density distributions of 
 galaxies within cluster virial radius of $r_{\mathrm{ctr}}<r_{\mathrm{200,cl}}$ and 
 galaxies in the cluster outskirts of $r_{\mathrm{ctr}}= 1\sim2 r_{\mathrm{200,cl}}$, respectively. 
 The thin contours enclose 0.5$\sigma$ and 1$\sigma$ of each sample, respectively.}
 \label{fig:fig3}
\end{figure}

Given the relatively low velocities of group galaxies \citep[about 50 to 400 $\kms$, ][]{Mcconnachie2009}, 
mergers and galaxy interactions are expected to be more frequent 
in group environments than in cluster environments,
so the intermediate-density region at $\rp<\rvir$ roughly corresponds to a group environment.
Figure~\ref{fig:fig3} below confirms that cluster member galaxies 
are mainly found in higher density region than roughly $\rtwenty\sim 20\barrho$
where more galaxies are gravitationally bound. 
The inner first and second grey contours in Figure~\ref{fig:fig2} 
include about 38\% and 68\% of the sample, respectively.  
Out of the sample, 17.5\% has a small pair separation of $\rp<\rvir$ and belongs to
groups or clusters.

%% New figure 2
To check where cluster member galaxies are actually distributed
in the $\rp$ and $\rho20$ space, we plot Figure~\ref{fig:fig3}.
By using the SDSS DR9 galaxy clusters catalogue of \citet[][]{Banerjee2018}, 
we found 68 galaxy clusters (with cluster richness $\Lambda_{200}>20$) at $0.045 \leq z <0.055$
overlapping the redshift range of our sample.
Only 15 clusters located within the survey region of our sample were finally selected.
Clusters sitting close to the survey boundaries were also excluded.

Based on the simulation result by \citet[][]{Serra2013}, 
we identified 131 spiral galaxies as cluster member galaxies 
from the galaxy sample with $0.042 \leq z <0.055$ and $\sigma > 70~\kms$.
Galaxies having a line-of-sight velocity difference from the clusters
below $1000\kms$ within clustercentric radii of $2r_{\mathrm{200,cl}}$ were conservatively selected.
The $r_\mathrm{200,cl}$ is the radius of the sphere 
whose mean overdensity drops to 200 times the critical density of the universe.
The mean $r_{\mathrm{200,cl}}$ of the 15 clusters is about 1~Mpc.
Out of 131 galaxies, 73 galaxies at $r_{\mathrm{ctr}}<r_{\mathrm{200,cl}}$
and 58 galaxies at $r_{\mathrm{ctr}}= 1\sim2 r_{\mathrm{200,cl}}$
were considered to be in the cluster virial and in-fall regions, respectively.
The orange and blue contours represent the number density distributions of the galaxies 
with $r_\mathrm{ctr}<r_{200,\mathrm{cl}}$ and $r_{\mathrm{ctr}}= 1\sim2 r_{\mathrm{200,cl}}$,
respectively. The 2D histogram represents the number density distribution
of the galaxy sample selected for this test. 
The contours for each sample enclose 0.5$\sigma$ and 1$\sigma$.

Figure~\ref{fig:fig3} shows that the identified cluster member galaxies 
mostly locate in the high-density region of $\rtwenty>10\barrho$.
For convenience, in this study, we define the region of $\rtwenty>20\barrho$ 
as a cluster environment.
Compared to the cluster galaxies in the cluster virial regon, 
galaxies in the cluster outskirts locate at relatively larger $\rp=1\sim2\rvir$.
When galaxies have smaller $\rp$ and larger $\rtwenty$, they are more likely to
locate at smaller clustercentric distances.

It can be seen that at smaller $\rp$s in the region of $\rtwenty>20\barrho$,
spiral galaxies suddenly disappear, which is a natural consequence,
given the morphology–density relation. 
That is, the smaller $\rp$ region where the spiral galaxies disappear 
corresponds to the inner region of clusters where the elliptical galaxy is preferentially located
In this context, given that more spiral galaxies are found in poor clusters (or groups) and often in their inner regions,
we can infer that many of galaxies with $\rp<0.5\rvir$ in a cluster environment may be 
members located in the inner region of poor clusters.

According to the hierarchical structure formation paradigm,
as $\rtwenty$ increases, galaxies join more and more rich and massive systems.
The region of $\rtwenty>50\barrho$ may contain many rich clusters.
The cluster catalogue we used seems to be missing many poor clusters. 
\citet[][]{Hwang2012} resolved the virialized cluster regions using the SDSS Abell clusters data
and studied AGN activity dependence of galaxies 
within $10r_{\mathrm{200,cl}}$ on the $r_{\mathrm{ctr}}$ and $\rp$.

We conclude that in order to adequately describe the environment of a galaxy,
one should consider the combined effect of the two $\rp$ and $\rtwenty$.

\subsubsection{Environmental dependence of AGN fraction in spirals with a high $\sigma$}
\label{sec:sec3.2.2}
\begin{figure*}
    \centering   
   \includegraphics[scale=0.9]{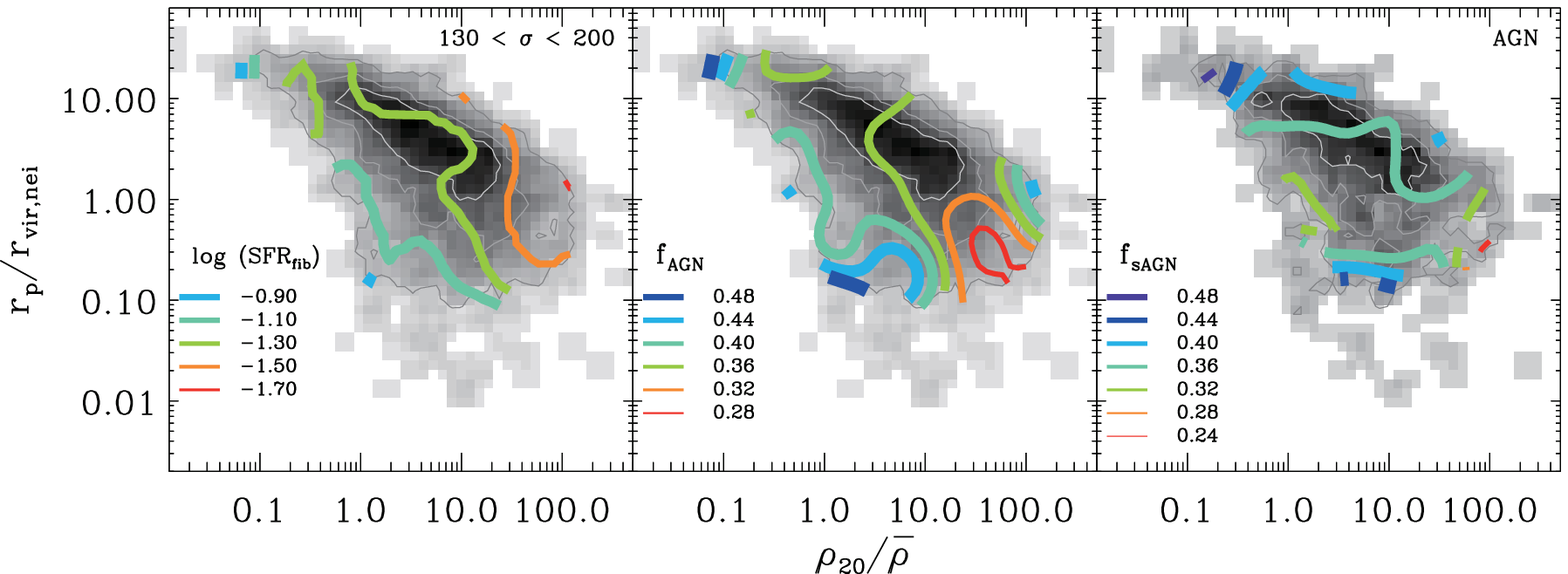}
    \caption{The distributions in the $\rp$–$\rtwenty$ plane of ${\rm log}~(\SFRf)$ (left) and $\fAGN$ (middle)
at given $\sigma$ with $130~\kms<\sigma<200~\kms$. 
The right-hand panel shows the distribution of the strong AGN fraction 
with log$~(L_{\rm [OIII]}/L_{\odot})>6.42$ in AGN host galaxies. 
The coloured thick contour lines denote each distribution.
The grey 2D histograms and thin contours represent 
the number density distributions of galaxies (left and middle) and AGN hosts (right), respectively.
The thin contours enclose 0.5$\sigma$ and 1$\sigma$ of each sample, respectively.}
    \label{fig:fig4}
\end{figure*}

\begin{figure*}
    \centering   
   \includegraphics[scale=0.9]{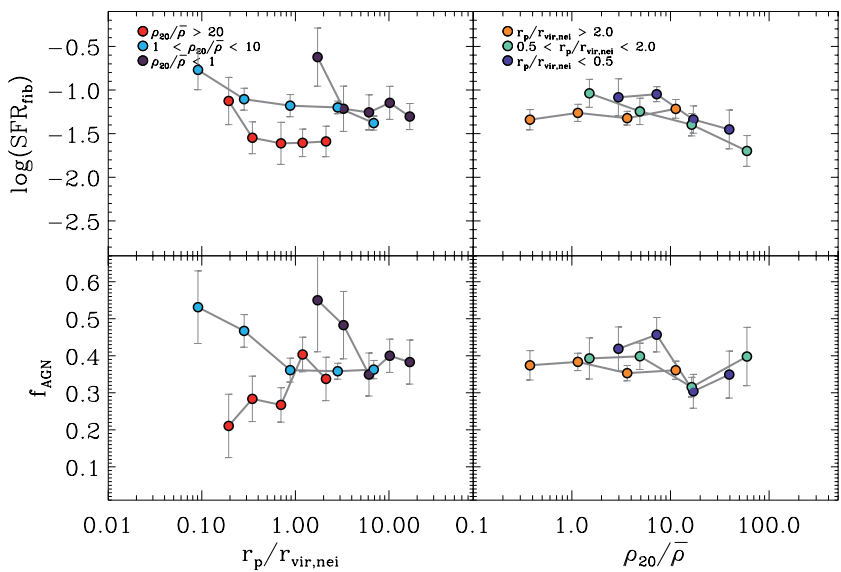}
    \caption{Behaviours of $\fagn$ and $\SFRfib$ as a function of $\rp$ at a fixed $\rtwenty$ (left)
    and as a function of $\rtwenty$ at a given $\rp$ (right) for the high-$\sigma$ sample.
    The colour of the dots changes depending on $\rtwenty$ (left) or $\rp$ (right). 
    The errors represent 1$\sigma$ deviation from 1000 bootstrap resampling.}
    \label{fig:fig5}
\end{figure*}

In Section~\ref{sec:sec3.1},
we have shown that AGN triggering primarily correlates with $\SFRf$ and $\sigma$ of galaxies.
In this section, by fixing those most influential properties,
we investigate the direct impact of the environment on AGN triggering.
First, we measure the impact for the $\sigma$-limited sample,
seen in the left and middle panels of Figure~\ref{fig:fig4}. 
To check the statistical significance of the several features found from Figure~\ref{fig:fig4},
we plot Figure~\ref{fig:fig5}.

%we measured f_AGN and SFR_fib as a function of rp at fixed rho_20 and as a function of rho_20 at fixed rp,

At a given $\rp$, $\SFRf$ increases with decreasing $\rtwenty$.
When $\rp<\rvir$, the $\rtwenty$ dependence of $\SFRf$ becomes obvious.
The highest $\SFRf$ enhancement is found in group galaxies with $\rp<0.3\rvir$,
implying that the neighbour interaction in groups is a dominant process of 
the central SF enhancement of a host galaxy.

On the other hand, interacting galaxies with $\rp < \rvir$ also undergo major changes in $\fagn$.
In particular, for group galaxies, the $\fagn$ rapidly increases at $\rp<0.3\rvir$ where the increase in $\SFRf$ occurs.
The luminosity of AGNs found there are brighter than those found elsewhere,
as seen from the right-hand panel of Figure~\ref{fig:fig4}. 
Several numerical simulations 
\citep{Mihos1996, Dimatteo2005, Springel2005, Hopkins2006, Debuhr2012} 
and observations \citep{Alonso2007, Ellison2011, Sabater2013}
have shown that the close pair interaction induces gas inflow into the galactic centre, 
enhancing both central SF and AGN activity, which is consistent with this finding.

Meanwhile, isolated galaxies experience less change in $\fagn$ and $\SFRf$ at a given $\rtwenty$, 
but show an apparent change in the $L_\mathrm{[OIII]}$. 
For example, for the most isolated galaxies with the largest $\rp$ at a given $\rtwenty$, 
AGNs are not detected more frequently but the detected AGNs tend to be more luminous. 

%The results are in good agreement with the observation 
%that the most luminous AGNs are related to major mergers \citep[][]{Treister2012}.
%whereas the the stochastic accretion in non-merging and secular process are dominant gas inflow mechanisms 
%in forming low and intermediate luminous AGNs \citep[see][]{Hopkins2014}. 

The most dramatic features are found in the cluster environment of $\rtwenty >20\barrho$,
as shown in Figures~\ref{fig:fig4} and \ref{fig:fig5}. 
The $\fagn$ of galaxies with $\rp < 0.5 \rvir$ decreases as $\rtwenty$ increases
and is the lowest at $\rtwenty \sim 50\barrho$.
On the other hand, 
despite the lowest $\SFRf$,
galaxies with $0.5\rvir < \rp < 2 \rvir$ (i.e. cluster outskirts) 
show a significant excess of the AGN fraction when $\rtwenty >50\barrho$ (e.g. in rich clusters),
but the level of AGN activity of them is weak.

\subsection{Different environmental effects depending on central SFR}
\label{sec:sec3.3}
In this section, to examine the direct environmental effect on the nuclear activity of galaxies,
we divide the high-$\sigma$ sample into two subsamples according to the $\SFRf$ values
as seen in Figure~\ref{fig:fig1}.
The high-$\SFRf$ subsample consists of 670 galaxies, 
corresponding to about 22.8\% of the high-$\sigma$ sample.
Table~\ref{tab:tab2} lists the statistics of the full sample and subsamples we use. 

\begin{table*}
   \centering
   \caption{Sample statistics}
   \label{tab:tab2}
   \begin{tabular}{lccr} % four columns, alignment for each
      \hline
      \hline      
       Number (Fraction) & Total & AGNs & $^{a}$Strong AGNs \\
      \hline
         All ({\footnotesize$\sigma > 70~\kms$}) & 11 096 (1.00) & 2942 (0.27) &  765 (0.07)\\
        \hline
         All ({\footnotesize$130~\kms<\sigma<200~\kms$}) &  2935 (1.00)  & 1079 (0.37) & 400 (0.14) \\ 
         ~~~~~~ High $\SFRf$ sample   &   670 (1.00) & 231 (0.35) & 158 (0.24) \\
         ~~~~~~~Low $\SFRf$ sample   &  2265 (1.00) & 848 (0.38) & 242 (0.11) \\
      \hline
      \multicolumn{3}{l}{Note: $^{a}$ Galaxies hosting a luminous AGN with $L_{\rm [OIII]} > 10^{40} \rm erg~s^{-1}$.}\\
   \end{tabular}
\end{table*}

\subsubsection{Galaxies with high central star formation rate}
\label{sec:sec3.3.1}

This subsample of galaxies consists of SFGs and starburst-AGN composite hosts.
In particular, in galaxies with a higher $\SFRf$, higher luminosity composite hosts are found but 
the number is much smaller than that of SFGs. 
We examine the distributions of $\SFRf$, $\fSFG$, and $\fagn$ of the sample in 
the $\rp$–$\rtwenty$ space.

\begin{figure*}
    \centering   
   \includegraphics[scale=0.9]{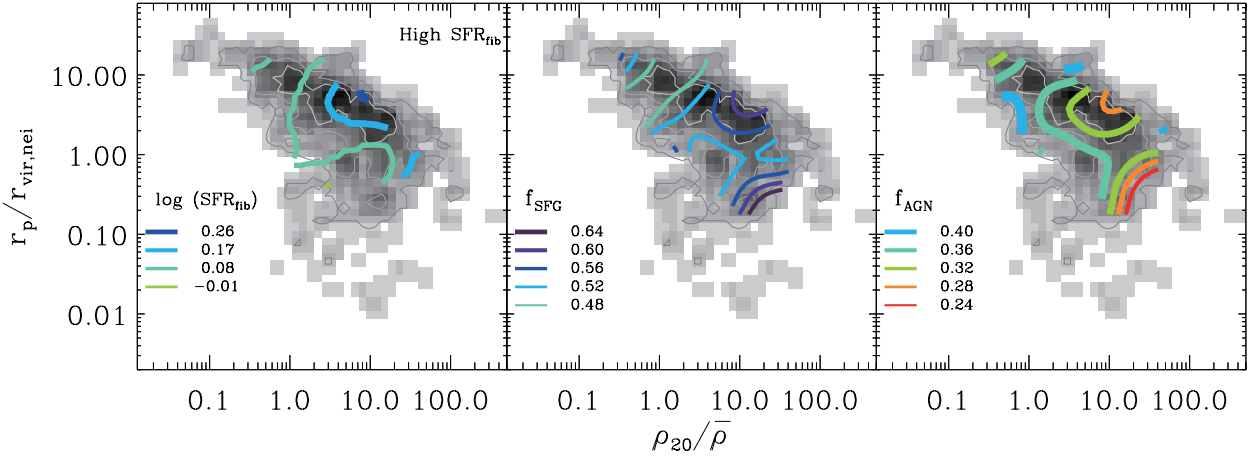}
    \caption{Distributions in the $\rp$–$\rtwenty$ plane of ${\rm log}~(\SFRf)$ (left), $\fSFG$ (middle), and $\fAGN$ (right) 
of the high-$\SFRf$ galaxy sample. For the log $(\SFRf)$ distribution, 
contours with a standard error of 1000 bootstrap estimates greater than 0.08 are eliminated.
The grey 2D histogram and the grey thin contours represent the number density distributions of the sample.}
    \label{fig:fig6}
\end{figure*}

The results are given in Figure~\ref{fig:fig6}.
The high-$\SFRf$ galaxies are hardly found in rich clusters.
On average, $\fSFG$ is larger than $\fagn$ over the entire space.
Table~\ref{tab:tab2} shows that the AGNs hosted by the high-$\SFRf$ galaxies 
are more likely to be powerful.
%
%We found interesting feature either 
%mostly isolated galaxies in intermediate-density regions 
%or galaxies in the cluster region. 

The most notable feature is that over the entire space,
the relation of an environment with $\fagn$ and $\fSFG$ is opposite overall.
We find that the SFGs are most often found in two specific environments
where the starburst-AGN composite hosts are least found.
One occupies a region of $\rp>>\rvir$ and $\rtwenty \sim 10\barrho$
where galaxies are expected to have recently undergone a major merger in a rich group.
The other occupies a region of $\rp<0.3\rvir$ and $7\barrho<\rtwenty<30\barrho$
which switches from group to cluster environment. 
Their $\fSFG$ and $\fagn$ values are comparable to each other.

For the latter environment, the galaxies no longer exist in a very dense environment of $\rtwenty>40\barrho$.
At $\rp<0.3\rvir$ and $\rtwenty <7\barrho$ (mostly groups), $\fSFG$ and $\fagn$ change little
but when $\rtwenty$ increases further, the values sensitively change,
requiring that physical mechanisms in the clusters should be more effective 
than the galaxy interactions in the groups.
As the system to which galaxies belong grows hierarchically, 
they are likely to experience a variety of physical mechanisms 
that funnel abundant gases to the centre.
Meanwhile, their $\fSFG$ and $\fagn$ values comparable
to those of galaxies at late-stage of the gas-rich mergers in groups
suggest that these galaxies may favour a merger-driven evolutionary scenario.

Figure~\ref{fig:fig6} also shows that
at those two locations, SFGs are found almost twice as many as AGN hosts.
Although elsewhere, less SFGs are found and more AGN hosts are found.
Interestingly, compared to the fractions in the two environments
it seems that the decrease in $\fSFG$ has led to the increase in $\fagn$ .
The result indicates that the violent environments, possibly associated with merger events,
cause a high gas accretion rate, not only inducing starbursts and rapid BH growth
but also obscuring AGNs by dense gas and dust \citep[see][, for a review]{Hickox2018}. 
On the other hand, other environments provide a moderate gas accretion rate, 
reducing $\fSFG$ and more AGNs becomes observable. 
In a field environment where galaxies do not belong to a group or cluster,
the lowest $\fSFG$ and the highest $\fagn$ are found.

These interesting features associated with starburst-AGN connection are unveiled
only after excluding the low-$\SFRf$ galaxies (77\% of the whole sample). 
This high-$\SFRf$ sample also demonstrates 
that an environment controls a gas accretion rate, eventually controlling 
a central SF and an obscuring medium nearby AGNs that intercepts AGN luminosity.
Consequently, it is difficult to observe the direct impact of the environment 
with this gas-rich sample.

\subsubsection{Galaxies with low central star formation rate}
\label{sec:sec3.3.2}
\begin{figure*}
    \centering   
   \includegraphics[scale=0.9]{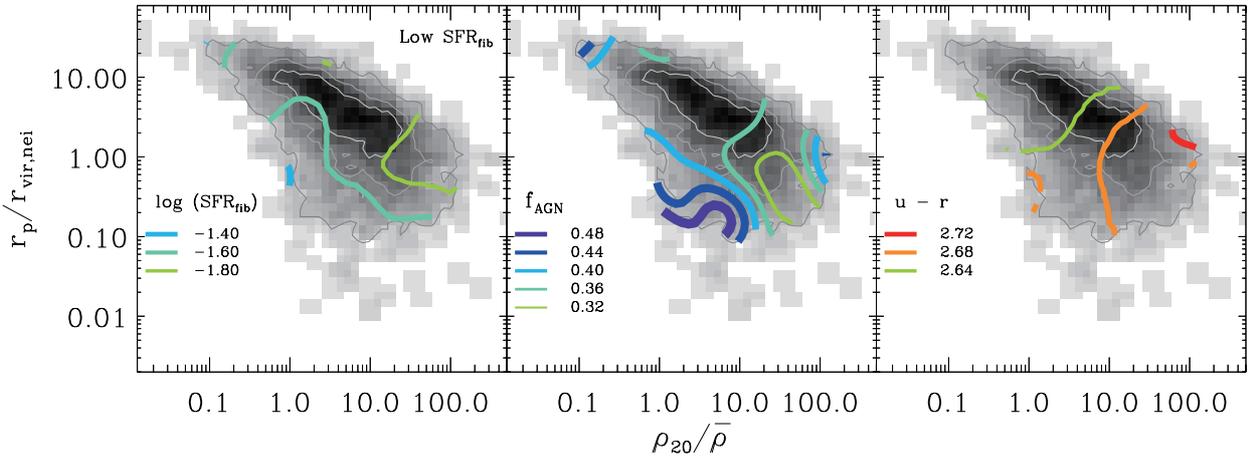}
    \caption{Distributions in the $\rp$–$\rtwenty$ plane 
of log $(\SFRf)$ (left), $\fAGN$ (middle), and $u-r$ colour (right) 
of the low-$\SFRf$ galaxy sample.
The grey 2D histogram and the grey thin contours show 
the number density distributions of the sample. 
For the log $(\SFRf)$ and $u-r$ distributions, 
contours with a standard error of 1000 bootstrap estimates greater than 0.1 and 0.02 are eliminated,
respectively.}
    \label{fig:fig7}
\end{figure*}

In this section, 
we examine the distributions in $\rp$-$\rtwenty$ space for $\SFRf$, $\fAGN$ and $u-r$ colour
of the low-$\SFRf$ sample. 
The results are shown in Figure~\ref{fig:fig7}.

\begin{figure*}
    \centering   
   \includegraphics[scale=0.9]{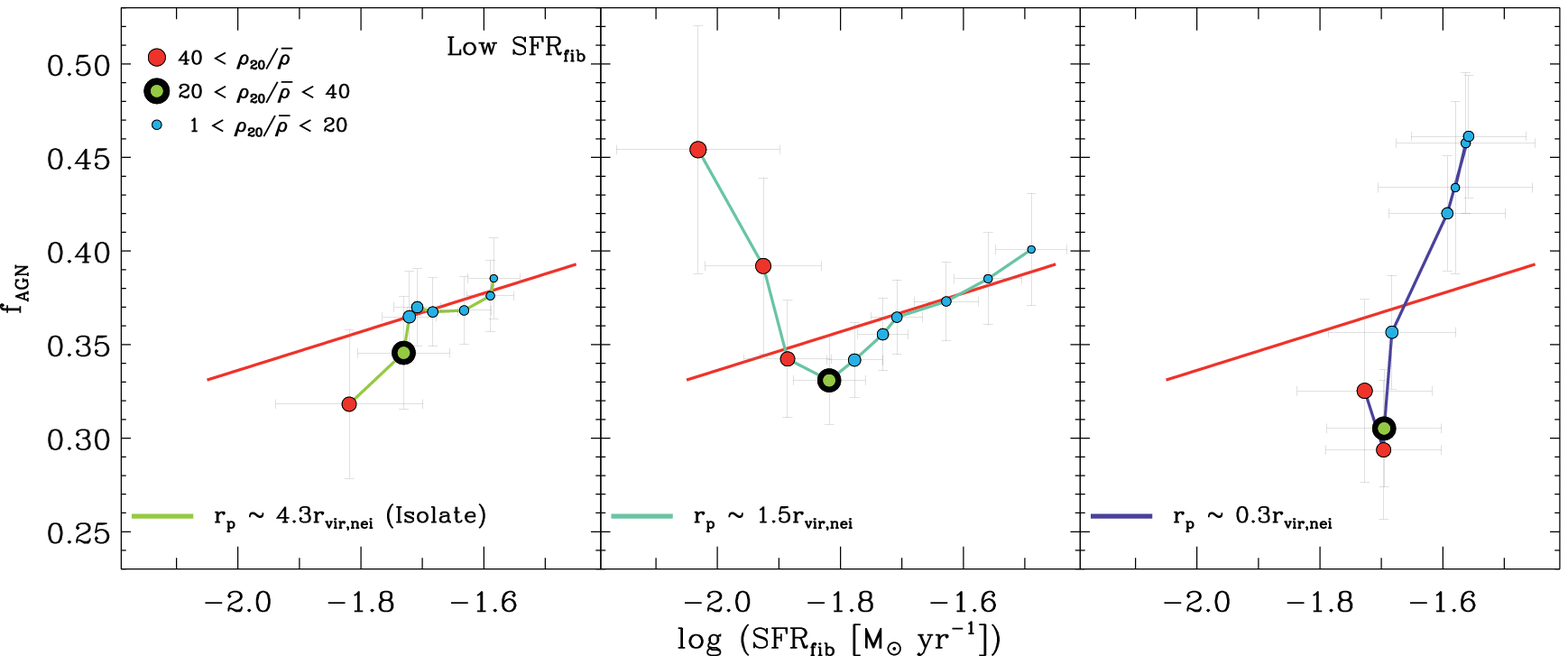}
    \caption{The $\rtwenty$ dependency of $\SFRf$ and $\fagn$ at given $\rp$ for the low-$\SFRfib$ sample.
Three cases with different $\rp$ values are investigated: (left)
isolated galaxies ($\rp \sim 4.3\rvir$), 
(middle) galaxies just about to enter their nearest neighbour's virial radius ($\rp \sim 1.5\rvir$),
and (right) galaxies in close interaction ($\rp \sim 0.3\rvir$).
The size of bins used to fix the $\rp$ and $\rtwenty$ values of the sample is 
$\Delta ({\rm log}{~\rp/\rvir})=0.23$ and $\Delta ({\rm log}{~\rtwenty/\barrho})=0.23$, respectively.
The colour and size of the dots change depending on $\rtwenty$. 
The errors represent 1-$\sigma$ deviation from 1000 bootstrap resampling.
For comparison, the relation between $\SFRf$ and $\fAGN$ of the isolated galaxies
is drawn in each panel with a red line.}
    \label{fig:fig8}
\end{figure*}

Since this $\SFRf$-limited subsample has slightly different $\SFRf$ values,
the $\SFRf$ distribution slightly depends on $\rp$ and $\rtwenty$.
Therefore, it is not easy to estimate the $\rp$ or $\rtwenty$ dependence of $\fagn$ at a given $\SFRf$
by comparing the left and middle panels of Figure~\ref{fig:fig7}.
Therefore, we plot Figure~\ref{fig:fig8} 
showing how $\fagn$ and $\SFRf$ vary depending on $\rtwenty$ at a given $\rp$.
We investigate three different $\rp$ cases: 
isolated galaxies ($\rp \sim 4.3 \rvir$), 
galaxies just entering the nearest neighbour's virial radius ($\rp \sim 1.5\rvir$), 
and close-interacting galaxies ($\rp\sim0.3\rvir$) are examined separately.
Each circle represents the median values for the galaxies at a fixed $\rtwenty$.
The error bars are calculated as bootstrapping errors using the bootstrapping method with 1000 runs.
%the rich cluster environment of $\rtwenty>40\barrho$ is marked with red dots.

Overall, galaxies in this subsample have low $\SFRf$, low AGN power, and red colour. 
For the isolated galaxy case, $\SFRf$ in groups slightly depends on $\rtwenty$
and $\fagn$ changes little.
They are almost free from the influence of the neighbours and background galaxies. 
Their $\SFRf$–$\fagn$ relation is used as a control and for comparison,
is drawn with a red line in each panel. 

Figure~\ref{fig:fig7} shows that the low-$SFRf$ sample nearly retains 
the diverse environmental dependences of $\fagn$ seen in the whole $\SFRf$ sample, 
particularly at $\rp<\rvir$. 
At $\rtwenty <10\barrho$ (i.e. groups), $\fagn$ is mostly sensitive to $\rp$.
The $\fagn$ rapidly increases at $\rp<0.3\rvir$ while $\SFRf$ changes little at $\rp<0.3\rvir$,
implying that at a given $\SFRf$, close neighbour interactions in groups 
are significantly effective in enhancing $\fagn$.

Meanwhile, the relation between $\fagn$ and $\SFRf$ 
in clusters becomes complex, completely different from that in groups.
At $\rtwenty \sim 50\barrho$ where $\fagn$ has the lowest value,
as $\rp$ decreases, $\fagn$ decreases but $\SFRf$ rather increases.
This trend is also seen in the whole sample (see Figs.~\ref{fig:fig4} and \ref{fig:fig5}).
As a galaxy moves within its cluster, a mechanism such as ram-pressure becomes dominant
\citep{Gunn1972}, likely compressing the gas into dense clouds, increasing $\SFRf$ but not $\fagn$.

At $\rp\sim0.3\rvir$, $\fagn$ significantly decreases as a galaxy joins more and more rich systems
while $\SFRf$ weakly decreases.
Figure~\ref{fig:fig7} also shows that
galaxies belonging to the outer region of rich clusters show an $\fagn$ enhancement, 
although they have the lowest $\SFRf$ and the reddest $u-r$ colour.
The green and red dots in the middle panel of Figure~\ref{fig:fig8} 
show that the result
is systematically significant although the errors are large.
The reddest $u-r$ colour in the outer region of massive clusters
implies that SF quenching has been already pre-processed.
Despite the lowest $\SFRf$ in the outskirts of rich clusters, the strong enhancement in $\fagn$
requires a violent mechanism.
\citet{Martini2009} found substantial numbers of AGNs on radial orbits in clusters and argued that they 
recently entered the clusters. Some studies found that
the AGN fractions of spiral galaxies at outer cluster radii are as large as those
of field galaxies \citep{Haggard2010, Hwang2012, Sohn2013}.

\section{Discussion}
\label{sec:sec4}

\subsection{High central SF and obscured AGN: the starburst-AGN connection}
\label{sec:sec4.1}
In general, if the gas causing starburst is also responsible for feeding the AGN, 
then a positive correlation is expected. However,
we identified galaxies with a negative correlation (high $\SFRf$ but low $\fagn$)
in two specific environments.  

The $\rp$–$\rtwenty$ diagram allows us to deduce what mechanisms cause 
large quantities of gas to inflow into the central regions, leading to rapid SF quenching.
In the region of $\rp>>\rvir$ and the intermediate-$\rtwenty$s,
the lowest $\fagn$ and the highest $\fSFG$ are found.
The AGNs observed at this stage (i.e. starburst-AGN composite)
are small in number but have high luminosity.
We infer that they are at late-stage of the gas-rich mergers inducing high accretion rates
 \citep[e.g.,][]{Park2009a},
which is in agreement with the observation
that the most luminous AGNs require major mergers \citep[][]{Treister2012}.
Here, we speculate that surrounded by dense gases and dust,
AGNs are hidden in SFGs (particularly, massive ones), leading to the lowest $\fagn$.
\citet[][]{Barnes1991} suggested in a simulation that violent tidal forces occurred during a
gas-rich galaxy merger trigger the bar formation supplying the gas fuel to a BH
and a violent SF in central gas clouds is accompanied. 
Indeed, infrared and hard X-ray-selected AGNs less sensitive to 
obscuration by gas and dust have apparent connections to galaxy mergers 
\citep[e.g.,][]{Veilleux2009, Koss2010, Satyapal2014, Kocevski2015, Ricci2017}.
\citet{Satyapal2014} found by using cross-matched SDSS and 
Wide-field Infrared Survey Explorer (WISE) data that 
galaxies in late-stage mergers show a higher excess of IR-selected AGNs than 
optically selected AGNs, even in post-merger galaxies sample. 

Surprisingly, the starburst-AGN connection-related feature is
found even in a dense environment.
As a gas-rich galaxy moves within clusters, 
repeated high-speed encounters \citep[or weak encounters,][]{Moore1998}
or tidals in clusters can disturb galaxies or direct gas inflow towards the central region, 
thus enhancing $\SFRf$ and triggering a lot of AGNs.
The interaction between galaxy and hot intra-cluster medium (ICM) directly can also affect AGN activity of cluster galaxies. 
For example, the ram-pressure stripping in clusters can also compress 
the galaxy interstellar medium, which also enhances the SFR 
\citep[e.g.,][]{Fujita1999}. 
\citet{Kronberger2008} found in simulated spiral galaxies 
that the SF enhancement can take place in the central part of galaxies, 
where the interstellar medium is compressed
by the ram-pressure of the surrounding ICM.
\citet{Bekki2003, Bekki2011} also demonstrated that ram-pressure induces the
collapse of molecular clouds and consequently trigger a starburst. 
\citet{Poggianti2017} have discovered a causal connection between
ram-pressure stripping and AGN activity from a jellyfish galaxy in clusters.
In contrast, the ram pressure at the centre of rich clusters
removes a significant amount of cold gas within a galaxy, leading to SF quenching.

As an alternative, the group-cluster merger is possible.
Gas-rich galaxies in in-falling group
may experience mergers and strong encounters in the in-fall region of the cluster
\citep[e.g.,][]{Vijayaragh2013}.
\citet{Hwang2018} showed in numerical simulation that 
while spiral galaxies fall into the cluster environment, 
galaxy–galaxy multiple interactions at high-speed enhance SF activity in spiral galaxies. 
\citet{Jaffe2016} also showed that gas-rich galaxies are found in the recent in-fall region
of the cluster or even inside the cluster.

In these violent environments, some of these may undergo a morphological transformation into 
gas-poor S0 galaxies \citep[e.g.][]{D'onofrio2015}. 
Unfortunately, we could not see perturbed signatures in the SFGs 
due to the spatial resolution of the SDSS imaging.
Along with these results, the reduction in $\fSFG$ and the enhancement in $\fagn$ 
measured in other relatively mild environments indicate that an environment affects 
a gas accretion rate, contributing to SF quenching and AGN detection,
particularly for gas-rich galaxies. 

\subsection{Enhanced AGN fraction in cluster environments}
\label{sec:sec4.2}
The most dramatic environmental impact is found
in quiescent spiral galaxies with relatively large pair separations ($0.5\rvir < \rp < 2 \rvir$)
in the very high-density region of $\rtwenty >50\barrho$. 
It is estimated that they are in the outskirts of rich clusters.
They have a higher $\fagn$ than in cluster centres and the AGNs have low luminosity. 
They also have rather redder colour and lower $\SFRf$ even when compared to galaxies inside the clusters,
suggesting that they have already experienced a `sudden' SF quenching mechanism. 

Galaxy starvation \citep{Larson1980} occurs on the cluster outskirts 
but is not suitable for the rapid SF quenching.
Once galaxies enter the cluster, in the beginning, 
tidal effects via the large gravitational potential of clusters can cause 
hot gas to escape within a galaxy, but not cold gas. 
Accordingly, the cold gas is gently removed through SF.
In contrast, ram-pressure stripping can remove 
a significant amount of cold gas within a galaxy, resulting in quenching.
However, it requires a dense ICM and a high relative velocity of a galaxy to the ICM,
which is dominant at the centre of rich clusters.

Meanwhile, according to the hierarchical structure formation paradigm, 
clusters grow via the in-fall of galaxies (in groups) along the filaments surrounding a cluster.
\citet{Mcgee2009} and \citet{DeLucia2012} argued  
that about $40\sim50$\% of galaxies in massive clusters has arrived as members of in-falling groups 
and the in-falling group galaxies could already be subjected to `pre-processing' in
group environments, prior to entering the cluster.
\citet{Fassbender2012} argued that 
galaxies with relatively lower velocity dispersions than those of central cluster galaxies
may result from the in-falling groups, thus raising the galaxy merger rate. 
\citet{Koulouridis2019} suggested galaxy mergers as
a dominant mechanism of the X-ray-detected AGN excess found in the outskirts of massive clusters at $z\sim1$.
\citet{Vazza2011} also demonstrated a high merger rate in the outskirts of massive clusters 
by using high-resolution cosmological simulations.

Meanwhile, cluster galaxies 
with $\rp < 0.3\rvir$ at $\rtwenty \sim 40 \barrho$ may move faster
in the inner regions of clusters where
mergers and interaction become less frequent and most encounters occur as high-speed fly-bys.
Due to the strong ram-pressure stripping, they may contain only very little cold gas.
For the non-merging quiescent case,
the stochastic accretion through secular processes becomes dominant 
and drives less luminous AGN activity \citep[see][]{Hopkins2014}.
Indeed, our result shows that 
they have the lowest $\fagn$ in the $\rp$-$\rtwenty$ space and the AGNs appear to be less luminous.

\section{SUMMARY}
\label{sec:sec5}

We identified 1 079 Type II AGNs (36.8\%) in a volume- and $\sigma$-limited sample 
of 2,935 spiral galaxies with $0.02 < z < 0.055$,  $M_r < -19.5$,  and
$130~\kms<\sigma<200~\kms$, selected from SDSS DR7, 
to determine a direct environmental effects on AGN activity. 
To define the environmental measurements, we used a volume-limited sample 
with $0.02 < z < 0.055$ and $M_r < -19.0$.
The combined impacts of the large-scale background density 
and interactions with the most influential neighbour are investigated.
Since the environment can affect both the central SF and AGN activity, 
the central $\SFRf$ of the sample is additionally limited 
to separate only the effect on AGN activity.
Two subsamples of before and after AGN peak in the $\SFRf$-$\sigma$ diagram
are used.

Our primary results are summarized as follows:

\begin{enumerate}
\item 
The AGN fraction is related to the central SFR depending on
the availability of central gas fuels.

At a given $\sigma$, galaxies with very high central SFRs have the lowest AGN fraction,
but the luminosity of the detected AGNs is the highest.
As the central SFR considerably decreases, the AGN fraction reaches its peak.
The luminosity of the AGNs at the peak is already significantly reduced. 
The result shows that the central SFR and AGN luminosity (BH accretion rate) are closely related.

\item
Galaxies with the highest $\SFRf$ but the lowest $\fagn$
are found in two violent environments of groups and (poor) clusters.
We argue here that the low $\fAGN$ value is
due to the difficulty in observing triggered AGNs in this optical survey
rather than the difficulty in inducing AGN activity itself
and that according the starburst-AGN connection, 
the SFGs optically obscured AGNs.

Indeed, they are found in rich groups where recent merger products are often found
and in clusters where 
the combined effect of ram pressure and frequent weak encounters dominates.
Considering the merger-AGN connection, 
the latter galaxies may also be greatly affected by a hierarchical merger
that occurs during system growth from group to cluster.

These gas-rich galaxies with massive bulge 
clearly shows that an environment controls a gas accretion rate, eventually controlling 
a central SF and an obscuring medium nearby AGNs that intercepts AGN luminosity.
 
\item
The observational evidence of the direct environmental effect is clearly found
in red spiral galaxies resided in the outskirts of rich clusters.
They show a higher AGN fraction despite the lack of gas 
(indicated by the lowest $\SFRf$ and the reddest colour).
This suggests that the SF quenching process has already completed and
that a strong galaxy interaction or a galaxy merging that occurs in the cluster outskirts  
accelerate the accretion of remained surrounding matter onto a BH.

\end{enumerate}

This study highlights how important mergers are in driving AGN activity and galaxy evolution.
In particular, 
galaxy–galaxy mergers in group environments and group-cluster mergers in cluster outskirts clearly demonstrate AGN-galaxy co-evolution. 
Indeed, the role of galaxy mergers in fuelling AGN has often been debated. 
In particular, because the correlation between mergers and detected AGN fraction is highly sensitive 
to obscuration by gas and dust, optical surveys have not clarified this correlation.
Although we use optically selected AGNs found in the local Universe, 
by considering a volume-limited spiral galaxy sample according to central velocity dispersion
and central SFR, we yield observational evidence with high reliability 
regarding the impact of merger processes on AGN triggering depending on the local background galaxy density. 
The use of such a large data set has led to a detailed and
robust statistical description of the environmental dependence of AGN activity.

\section*{Acknowledgements}

We thank the anonymous referee for useful and detailed comments that improved significantly the original manuscript. 

We acknowledge support from the National Research Foundation (NRF) of 
Korea to the Center for Galaxy Evolution Research (No. 2017R1A5A1070354).  
The work by SSK was supported by the NRF grant 
funded by the Ministry of Science and ICT of Korea (NRF-2014R1A2A1A11052367). 
This work was also supported by the BK21 plus program through the NRF 
funded by the Ministry of Education of Korea. 

Funding for the SDSS and SDSS-II has been provided by the Alfred P. Sloan Foundation, the Participating Institutions, the National Science Foundation, the U.S. Department of Energy, the National Aeronautics and Space Administration, the Japanese Monbukagakusho, the Max Planck Society, and the Higher Education Funding Council for England. The SDSS Web site is http://www.sdss.org/.
The SDSS is managed by the Astrophysical Research Consortium for the Participating Institutions. The Participating Institutions are the American Museum of Natural History, Astrophysical Institute Potsdam, University of Basel, Cambridge University, Case Western Reserve University, University of Chicago, Drexel University, Fermilab, the Institute for Advanced Study, the Japan Participation Group, Johns Hopkins University, the Joint Institute for Nuclear Astrophysics, the Kavli Institute for Particle Astrophysics and Cosmology, the Korean Scientist Group, the Chinese Academy of Sciences (LAMOST), Los Alamos National Laboratory, the Max-Planck-Institute for Astronomy (MPIA), the Max-Planck-Institute for Astrophysics (MPA), New Mexico State University, Ohio State University, University of Pittsburgh, University of Portsmouth, Princeton University, the United States Naval Observatory, and the University of Washington.

%%%%%%%%%%%%%%%%%%%%%%%%%%%%%%%%%%%%%%%%%%%%%%%%%%



%\bibliographystyle{mnras}

\begin{thebibliography}{99}
\bibitem[\protect\citeauthoryear{Abazajian et al.}{2009}]{Abazajian2009}
 Abazajian, K.~N., et al., 2009, \apjs, 182, 543
\bibitem[\protect\citeauthoryear{Aird et al.}{2012}]{Aird2012}
 Aird, J., et al., 2012, \apj, 746, 90
\bibitem[\protect\citeauthoryear{Alexander \& Hickox}{2012}]{Alexander2012}
 Alexander, D.~M., \& Hickox, R.~C., 2012, New Astron. Rev., 56, 93
\bibitem[\protect\citeauthoryear{Alonso et al.}{2007}]{Alonso2007}
 Alonso, M.~S., Lambas, D.~G., Tissera, P., \& Coldwell, G., 2007, \mnras, 375, 1017
\bibitem[\protect\citeauthoryear{Alonso et al.}{2013}]{Alonso2013}
 Alonso, M.~S., Coldwell, G., \& Lambas, D.~G., 2013, \aap, 549, A141
\bibitem[\protect\citeauthoryear{Argudo-Fern\'{a}ndez et al.}{2016}]{Argudo2016}
 Argudo-Fern\'{a}ndez, M., Shen, S., Sabater, J., Duarte Puertas, S., Verley, S., Yang, X., 2016, \ana, 592, A30
\bibitem[\protect\citeauthoryear{Argudo-Fern\'{a}ndez et al.}{2018}]{Argudo2018}
 Argudo-Fern\'{a}ndez, M., Lacerna, I., \& Duarte Puertas, S., 2018, \ana, 620, A113
\bibitem[\protect\citeauthoryear{Baldwin et al.}{1981}]{Baldwin1981}
 Baldwin, J.~A., Phillips, M.~M., \& Terlevich, R., 1981, \pasp, 93, 5
\bibitem[\protect\citeauthoryear{Banerjee et al.}{2018}]{Banerjee2018}
 Banerjee, P., Szabo, T., Pierpaoli, E., Franco, G., Ortiz, M., Ormas, A., Tornello, B., 2018, New Astron., 58, 61
\bibitem[\protect\citeauthoryear{Barnes \& Hernquist}{1991}]{Barnes1991}
 Barnes, J.~E., Hernquist, L.~E., 1991, \apj, 370, L65
\bibitem[\protect\citeauthoryear{Batiste et al.}{2017}]{Batiste2017}
 Batiste, M., Bentz, M.~C., Raimundo, S.~I., Vestergaard, M., \& Onken, C.~A., 2017, \apj, 838, L10
\bibitem[\protect\citeauthoryear{Bekki \& Couch}{2003}]{Bekki2003}
 Bekki, K., \& Couch, W.~J., 2003, \apj, 596, L13
\bibitem[\protect\citeauthoryear{Bekki \& Couch}{2011}]{Bekki2011}
 Bekki, K., \& Couch, W.~J., 2011, \mnras, 415, 1783 
\bibitem[\protect\citeauthoryear{Bernardi et al.}{2003}]{Bernardi2003}
 Bernardi, M., et al., 2003, \aj, 125, 1817
\bibitem[\protect\citeauthoryear{Blanton et al.}{2005}]{Blanton2005}
 Blanton, M.~R., et al., 2005, \aj, 129, 2562
\bibitem[\protect\citeauthoryear{Brinchmann et al.}{2004}]{Brinchmann2004}
 Brinchmann, J., Charlot, S., White, S.~D.~M., Tremonti, C., Kauffmann, G., Heckman, T., Brinkmann, J., 2004, \mnras, 351, 1151
\bibitem[\protect\citeauthoryear{Chang et al.}{2017}]{Chang2017}
 Chang, Y.-Y., et al., 2017, \apjs, 233, 19
\bibitem[\protect\citeauthoryear{Cheung et al.}{2015}]{Cheung2015}
 Cheung, E., et al., 2015, \apj, 807, 36
\bibitem[\protect\citeauthoryear{Choi et al.}{2009}]{Choi2009}
 Choi, Y.-Y., Woo, J.-H., \& Park, C., 2009, \apj, 699, 1679
\bibitem[\protect\citeauthoryear{Choi et al.}{2010}]{Choi2010}
 Choi, Y.-Y., Han, D.-H., \& Kim, S.~S., 2010, J. Korean Astron. Soc., 43, 191
\bibitem[\protect\citeauthoryear{Cid Fernandes et al.}{2010}]{Cidfernandes2010}
 Cid Fernandes, R., et al., 2010, \mnras, 403, 1036
\bibitem[\protect\citeauthoryear{Cid Fernandes et al.}{2011}]{Cidfernandes2011}
 Cid Fernandes, R., Stasi\'{n}ska, G., Mateus, A., \& Vale Asari, N., 2011, \mnras, 413, 1687
\bibitem[\protect\citeauthoryear{Ciotti \& Ostriker}{2007}]{Ciotti2007}
 Ciotti, L., \& Ostriker, J.~P., 2007, \apj, 665, 1038
\bibitem[\protect\citeauthoryear{Combes}{2003}]{Combes2003}
 Combes, F., 2003, in Collin S., Combes F., Shlosman I., eds, ASP Conf. Ser. Vol. 290, Active Galactic Nuclei: From Central Engine to Host Galaxy, Astron. Soc. Pac., San Francisco, p. 411
\bibitem[\protect\citeauthoryear{D'Onofrio et al.}{2015}]{D'onofrio2015}
 D'Onofrio, M., Marziani, P., \& Buson, L., 2015, Front. Astron. Space Sci., 2, 4
\bibitem[\protect\citeauthoryear{Darg et al.}{2010}]{Darg2010}
 Darg, D.~W., et al., 2010, \mnras, 401, 1552
\bibitem[\protect\citeauthoryear{Davies et al.}{2012}]{Davies2012}
 Davies, R., Burtscher, L., Dodds-Eden, K., \& Orban de Xivry, G., 2012, J. Phys. Conf. Ser. 372, 012046
\bibitem[\protect\citeauthoryear{DeBuhr et al.}{2012}]{Debuhr2012}
 DeBuhr, J., Quataert, E., \& Ma, C.-P., 2012, \mnras, 420, 2221
\bibitem[\protect\citeauthoryear{De Lucia et al.}{2012}]{DeLucia2012}
 De Lucia, G., Weinmann, S., Poggianti, B.~M., Arag\'{o}n-Salamanca, A., \& Zaritsky, D., 2012, \mnras, 423, 1277
\bibitem[\protect\citeauthoryear{Di Matteo et al.}{2005}]{Dimatteo2005}
 Di Matteo, T., Springel, V., \& Hernquist, L., 2005, \nat, 433, 604
\bibitem[\protect\citeauthoryear{Ellison et al.}{2011}]{Ellison2011}
 Ellison, S.~L., Patton, D.~R., Mendel, J.~T., \& Scudder, J.~M., 2011, \mnras, 418, 2043
\bibitem[\protect\citeauthoryear{Elmegreen et al.}{1998}]{Elmegreen1998}
 Elmegreen, B.~G., et al., 1998, \apj, 503, L119
\bibitem[\protect\citeauthoryear{Fassbender et al.}{2012}]{Fassbender2012}
 Fassbender, R., \v{S}uhada, R., \& Nastasi, A., 2012, Adv. Astron., 2012, 138380
\bibitem[\protect\citeauthoryear{Ferrarese \& Merritt}{2000}]{Ferrarese2000}
 Ferrarese, L., \& Merritt, D., 2000, \apj, 539, L9
\bibitem[\protect\citeauthoryear{Fisher \& Drory}{2010}]{Fisher2010}
 Fisher, D.~B., \& Drory, N., 2010, \apj, 716, 942
\bibitem[\protect\citeauthoryear{Fujita \& Nagashima}{1999}]{Fujita1999}
 Fujita, Y., Nagashima, M., 1999, \apj, 516, 619
\bibitem[\protect\citeauthoryear{Gebhardt et al.}{2000}]{Gebhardt2000}
 Gebhardt, K., et al., 2000, \apj, 539, L13
\bibitem[\protect\citeauthoryear{Goulding et al.}{2018}]{Goulding2018}
 Goulding, A.~D., et al., 2018, \pasj, 70, S37
\bibitem[\protect\citeauthoryear{Grogin et al.}{2005}]{Grogin2005}
 Grogin, N.~A., et al., 2005, \apj, 627, L97
\bibitem[\protect\citeauthoryear{G\"{u}ltekin et al.}{2009}]{Gultekin2009}
 G\"{u}ltekin, K., et al., 2009, \apj, 698, 198
\bibitem[\protect\citeauthoryear{Gunn \& Gott}{1972}]{Gunn1972}
 Gunn, J.~E., \& Gott, J.~R., 1972, \apj, 176, 1
\bibitem[\protect\citeauthoryear{Haan et al.}{2009}]{Hann2009}
 Haan, S., Schinnerer, E., Emsellem, E., Garcia-Burillo, S., Combes, F., Mundell, C.~G., Rix, H.-W., 2009, \apj, 692, 1623
\bibitem[\protect\citeauthoryear{Haggard et al.}{2010}]{Haggard2010}
 Haggard, D., Green, P.~J., Anderson, S.~F., Constantin, A., Aldcroft, T.~L., Kim, D.-W., Barkhouse, W.~A., 2010, \apj, 723, 1447
\bibitem[\protect\citeauthoryear{Hernquist \& Mihos}{1995}]{Hernquist1995}
 Hernquist, L., \& Mihos, J.~C., 1995, \apj, 448, 41
\bibitem[\protect\citeauthoryear{Hickox \& Alexander}{2018}]{Hickox2018}
 Hickox, R.~C., \& Alexander, D.~M., 2018, \araa, 56, 625
\bibitem[\protect\citeauthoryear{Hopkins et al.}{2005}]{Hopkins2005}
 Hopkins P. F., Hernquist L., Cox T. J., Di Matteo, T., Martini, P., Robertson, B., Springel, V., 2005, \apj, 630, 705
\bibitem[\protect\citeauthoryear{Hopkins et al.x}{2006}]{Hopkins2006}
 Hopkins, P.~F., Hernquist, L., \& Cox, T.~J., 2006, \apjs, 163, 1
\bibitem[\protect\citeauthoryear{Hopkins et al.}{2008}]{Hopkins2008}
 Hopkins, P.~F., Hernquist, L., Cox, T.~J., \& Kere\v{s}, D., 2008, \apjs, 175, 356
\bibitem[\protect\citeauthoryear{Hopkins et al.}{2014}]{Hopkins2014}
 Hopkins, P.~F., Kocevski, D.~D., \& Bundy, K., 2014, \mnras, 445, 823
\bibitem[\protect\citeauthoryear{Hwang et al.}{2012}]{Hwang2012}
 Hwang, H.~S., Park, C., Elbaz, D., \& Choi, Y.-Y., 2012, \aap, 538, A15
\bibitem[\protect\citeauthoryear{Hwang et al.}{2018}]{Hwang2018}
 Hwang, J.-S., Park, C., Banerjee, A., \& Hwang, H.~S., 2018, \apj, 856, 160
\bibitem[\protect\citeauthoryear{Ishibashi \& Fabian}{2016}]{Ishibashi2016}
 Ishibashi, W., \& Fabian, A. C., 2016, \mnras, 463, 1291
\bibitem[\protect\citeauthoryear{Jaff\'{e} et al.}{2016}]{Jaffe2016}
 Jaff\'{e}, Y. L. et al., 2016, \mnras, 461, 1202
\bibitem[\protect\citeauthoryear{Kauffmann et al.}{2003}]{Kauffmann2003}
 Kauffmann, G., et al., 2003, \mnras, 346, 1055
\bibitem[\protect\citeauthoryear{Kauffmann et al.}{2007}]{Kauffmann2007}
 Kauffmann, G., et al., 2007, \apjs, 173, 357
\bibitem[\protect\citeauthoryear{Kewley et al.}{2001}]{Kewley2001}
 Kewley, L.~J., Dopita, M.~A., Sutherland, R.~S., Heisler, C.~A., \& Trevena, J., 2001, \apj, 556, 121
\bibitem[\protect\citeauthoryear{Kewley et al.}{2006}]{Kewley2006}
 Kewley, L.~J., Groves, B., Kauffmann, G., \& Heckman, T., 2006, \mnras, 372, 961
\bibitem[\protect\citeauthoryear{Kim et al.}{2019}]{Kiminprep}
 Kim, M., Choi, Y.-Y., \& Kim, S.~S., \mnras, in submitted
\bibitem[\protect\citeauthoryear{Kocevski et al.}{2015}]{Kocevski2015}
 Kocevski D. D., et al., 2015, \apj, 814, 104
\bibitem[\protect\citeauthoryear{Kormendy \& Kennicutt}{2004}]{Kormendy2004}
 Kormendy, J., \& Kennicutt, Jr.~R.~C., 2004, \araa, 42, 603
\bibitem[\protect\citeauthoryear{Koss et al.}{2010}]{Koss2010}
 Koss M., Mushotzky R., Veilleux S., Winter L., 2010, ApJL, 716, L125
\bibitem[\protect\citeauthoryear{Koulouridis \& Bartalucci}{2019}]{Koulouridis2019} 
 Koulouridis, E., \& Bartalucci, I., 2019, \ana, 623, L10
\bibitem[\protect\citeauthoryear{Kronberger et al.}{2008}]{Kronberger2008}
 Kronberger T., Kapferer W., Ferrari C., Unterguggenberger S., \& Schindler, S., 2008, \ana, 481, 337
\bibitem[\protect\citeauthoryear{LaMassa et al.}{2013}]{Lamassa2013} 
 LaMassa, S.~M., Heckman, T.~M., Ptak, A., \& Urry, C.~M., 2013, \apj, 765, L33
\bibitem[\protect\citeauthoryear{Larson et al.}{1980}]{Larson1980} 
 Larson, R.~B., Tinsley, B.~M., \& Caldwell, C.~N., 1980, \apj, 237, 692
\bibitem[\protect\citeauthoryear{Liu et al.}{2012}]{Liu2012} 
 Liu, X., Shen, Y., \& Strauss, M.~A.\ 2012, \apj, 745, 94
\bibitem[\protect\citeauthoryear{Lynden-Bell}{1969}]{Lyndenbell1969} 
 Lynden-Bell, D., 1969, \nat, 223, 690
\bibitem[\protect\citeauthoryear{Martini}{2009}]{Martini2009} 
 Martini, P., Sivakoff, G.~R., \& Mulchaey, J.~S., 2009, \apj, 701, 66
\bibitem[\protect\citeauthoryear{McConnachie et al.}{2009}]{Mcconnachie2009} 
 McConnachie, A.~W., Patton, D.~R., Ellison, S.~L., \& Simard, L., 2009, \mnras, 395, 255
 \bibitem[\protect\citeauthoryear{McGee et al.}{2009}]{Mcgee2009} 
 McGee, S.~L., Balogh, M.~L., Bower, R.~G., Font, A.~S., \& McCarthy, I.~G., 2009, \mnras, 400, 937
\bibitem[\protect\citeauthoryear{Melnick \& De Propris}{2013}]{Melnick2013} 
 Melnick, J., \& De Propris, R., 2013, \mnras, 431, 2034
\bibitem[\protect\citeauthoryear{Mihos \& Hernquist}{1996}]{Mihos1996} 
 Mihos, J.~C., \& Hernquist, L., 1996, \apj, 464, 641
\bibitem[\protect\citeauthoryear{Moles et al.}{1995}]{Moles1995} 
 Moles, M., M\'{a}rquez, I., \& P\'{e}rez, E., 1995, \apj, 438, 604
\bibitem[\protect\citeauthoryear{Monaghan \& Lattanzio}{1985}]{Monaghan1985} 
 Monaghan, J.~J., \& Lattanzio, J.~C.\ 1985, \aap, 149, 135
\bibitem[\protect\citeauthoryear{Moore et al.}{1998}]{Moore1998} 
 Moore, B., Lake, G., \& Katz, N., 1998, \apj, 495, 139
\bibitem[\protect\citeauthoryear{Park \& Choi}{2005}]{Park2005} 
 Park, C., \& Choi, Y.-Y., 2005, \apj, 635, L29
\bibitem[\protect\citeauthoryear{Park et al.}{2007}]{Park2007} 
 Park, C., Choi, Y.-Y., Vogeley, M.~S., Gott, J.~R. III, Blanton, M.~R., SDSS Collaboration, 2007, \apj, 658, 898
\bibitem[\protect\citeauthoryear{Park et al.}{2008}]{Park2008} 
 Park, C., Gott, J.~R., \& Choi, Y.~Y., 2008, \apj, 674, 784
\bibitem[\protect\citeauthoryear{Park \& Choi}{2009}]{Park2009a} 
 Park, C., \& Choi, Y.-Y., 2009, \apj, 691, 1828
\bibitem[\protect\citeauthoryear{Poggianti et al.}{2017}]{Poggianti2017} 
 Poggianti, B.~M., Jaff\'{e}, Y.~L., Moretti, A., et al., 2017, \nat, 548, 304
\bibitem[\protect\citeauthoryear{Rees}{1984}]{Rees1984} 
 Rees, M.~J., 1984, \araa, 22, 471
\bibitem[\protect\citeauthoryear{Ricci et al.}{2017}]{Ricci2017} 
 Ricci C., et al., 2017,\mnras, 468, 1273
\bibitem[\protect\citeauthoryear{Sabater et al.}{2012}]{Sabater2012} 
 Sabater, J., Verdes-Montenegro, L., Leon, S., Best, P., \& Sulentic, J., 2012, \ana, 545, A15
\bibitem[\protect\citeauthoryear{Sabater et al.}{2013}]{Sabater2013} 
 Sabater, J., Best, P.~N., \& Argudo-Fern\'{a}ndez, M., 2013, \mnras, 430, 638
\bibitem[\protect\citeauthoryear{Sabater et al.}{2015}]{Sabater2015} 
 Sabater, J., Best, P.~N., \& Heckman, T.~M., 2015, \mnras, 447, 110
\bibitem[\protect\citeauthoryear{Salim et al.}{2007}]{Salim2007} 
 Salim, S., et al., 2007, \apjs, 173, 267
\bibitem[\protect\citeauthoryear{Sanders et al.}{1988}]{Sanders1988} 
 Sanders, D.~B., Soifer, B.~T., Elias, J.~H., Madore, B.~F., Matthews, K., Neugebauer, G., Scoville, N.~z., 1988, \apj, 325, 74
\bibitem[\protect\citeauthoryear{Satyapal et al.}{2014}]{Satyapal2014} 
 Satyapal, S., Ellison, S.~L., McAlpine, W., Hickox, R.~C., Patton, D.~R., Mendel, J.~T., 2014, \mnras, 441, 1297
\bibitem[\protect\citeauthoryear{Schawinski et al.}{2010}]{Schawinski2010} 
 Schawinski, K., et al., 2010, \apj, 711, 284
\bibitem[\protect\citeauthoryear{Scott \& Kaviraj}{2014}]{Scott2014} 
 Scott, C., \& Kaviraj, S., 2014, \mnras, 437, 2137
\bibitem[\protect\citeauthoryear{Serra \& Diaferio}{2013}]{Serra2013} 
 Serra, A.~L., \& Diaferio, A., 2013, \apj, 768, 116
\bibitem[\protect\citeauthoryear{Shlosman et al.}{1990}]{Shlosman1990} 
 Shlosman, I., Begelman, M.~C., \& Frank, J., 1990, \nat, 345, 679
\bibitem[\protect\citeauthoryear{Slavcheva-Mihova \& Mihov}{2011}]{Slav2011} 
 Slavcheva-Mihova, L., \& Mihov, B., 2011, \aap, 526, A43
\bibitem[\protect\citeauthoryear{Sohn et al.}{2013}]{Sohn2013} 
 Sohn, J., Hwang, H.~S., Lee, M.~G., Lee, G.-H., \& Lee, J.~C., 2013, \apj, 771, 106
\bibitem[\protect\citeauthoryear{Springel et al.}{2005}]{Springel2005} 
 Springel, V., Di Matteo, T., \& Hernquist, L., 2005, \apj, 620, L79
\bibitem[\protect\citeauthoryear{Stasi\'{n}ska et al.}{2008}]{Stasinska2008} 
 Stasi\'{n}ska, G., et al., 2008, \mnras, 391, L29
\bibitem[\protect\citeauthoryear{Treister et al.}{2012}]{Treister2012} 
 Treister, E., Schawinski, K., Urry, C.~M., \& Simmons, B.~D., 2012, ApJL, 758, L39
\bibitem[\protect\citeauthoryear{Tremaine et al.}{2002}]{Tremaine2002} 
 Tremaine, S., et al., 2002, \apj, 574, 740
\bibitem[\protect\citeauthoryear{Vazza et al.}{2011}]{Vazza2011} 
 Vazza, F., Roncarelli, M., Ettori, S., \& Dolag, K., 2011, \mnras, 413, 2305
\bibitem[\protect\citeauthoryear{Veilleux \& Osterbrock}{1987}]{Veilleux1987} 
 Veilleux, S., \& Osterbrock, D.~E., 1987, \apjs, 63, 295
\bibitem[\protect\citeauthoryear{Veilleux et al.}{2009}]{Veilleux2009} 
 Veilleux, S., et al., 2009, \apjs, 182, 628
\bibitem[\protect\citeauthoryear{Vijayaraghavan \& Ricker}{2013}]{Vijayaragh2013}
 Vijayaraghavan, R., \& Ricker, P.~M. 2013, \mnras, 435, 2713
\bibitem[\protect\citeauthoryear{Villforth et al.}{2014}]{Vill2014} 
 Villforth, C., et al., 2014, \mnras, 439, 3342
 \bibitem[\protect\citeauthoryear{von der Linden et al.}{2010}]{vonderlinden2010} 
 von der Linden, A., Wild, V., Kauffmann, G., White, S.~D.~M., \& Weinmann, S., 2010, \mnras, 404, 1231
\bibitem[\protect\citeauthoryear{Wada}{2004}]{Wada2004} 
 Wada, K., 2004, Coevolution of Black Holes and Galaxies ed. L. C. Ho (Cambridge: Cambridge Univ. Press), 186
\bibitem[\protect\citeauthoryear{Wada et al.}{2009}]{Wada2009} 
 Wada, K., Papadopoulos, P.~P., \& Spaans, M., 2009, \apj, 702, 63
\bibitem[\protect\citeauthoryear{Weston et al.}{2017}]{Weston2017} 
 Weston, M.~E., McIntosh, D.~H., Brodwin, M., Mann, J., Cooper, A., McConnel, A., Nielsen, J.~L., 2017, \mnras, 464, 3882
\end{thebibliography}


%%%%%%%%%%%%%%%%%%%%%%%%%%%%%%%%%%%%%%%%%%%%%%%%%%

%%%%%%%%%%%%%%%%%%%%%%%%%%%%%%%%%%%%%%%%%%%%%%%%%%

% Don't change these lines
\bsp	% typesetting comment
\label{lastpage}

\end{document}